\documentclass[11pt]{article}
\begin{document}
\title{\bf Perfect Spin Fluid with Intrinsic Color Charge}
\author{O. V. Babourova,\thanks{E-mail:baburova@orc.ru}\\
Department of Theoretical Physics, Faculty of Physics,\\
Moscow State University, Leninskie Gory, d.1, st.2, Moscow 119992, Russia,\\
A. S. Vshivtsev,\thanks{Deceased}\\
Academician V. P. Myasnikov,\\
Keldysh Institute of Applied Mathematics, Russian Academy of Sciences,\\
Miusskaya pl. 4, Moscow 125045, Russia,\\
\and B. N. Frolov,\thanks{E-mail:frolovbn@orc.ru}\\
Department of Mathematics, Moscow State Pedagogical University,\\
Krasnoprudnaya 14, Moscow 107140, Russia
}
\date{}
\maketitle
\vskip 0.4cm
\par
{\bf Abstract—--}{\small A variational theory of a perfect spin fluid with intrinsic
non-Abelian color charge is constructed with allowance for spin-polarization chromomagnetic effects
in Rie\-mann--Cartan space with curvature and torsion. The spacelike nature of the spin is taken into
account explicitly in this theory by including the Frenkel condition in the Lagrangian. The equations of motion,
the laws that govern the evolutions of the spin and color-charge tensors, and the expression for the energy-momentum
tensor for the fluid in question are obtained. In the limiting case, the theory goes over to the well-known
theory of Weyssenhoff--Raabe perfect spin fluid.}
\vskip 0.4cm
\par

\section{INTRODUCTION}
    The commonly accepted approach to the description of elementary particles relies on quantum field theory.
In this approach, a quantum field is treated as an infinite set of interacting oscillators. Thereby, a complicated
field-theoretical construction is approximated by a mechanical medium whose properties can be studied by applying
the existing mathematical formalism. It should be noted that one more mechanical model, based on the hydrodynamic
description of a medium \cite{Sed}--\cite{Wil}, underlies quantum field theory. The motivations for invoking this
approach are diverse. On one hand, it relies on the well-advanced mathematical formalism capable of taking into
account nontrivial topologies of gauge fields \cite{Mon}. On the other hand, the approach in question displayed
an astonishingly high predictive power both in QED and in elementary-particle theory, including the Landau
hydrodynamic model of multi-particle hadron production and the possibility of describing various phases
of quark-gluon plasma on the basis of fluid dynamics \cite{VP}. We do not dwell here on the theory of gravity,
where this approach is widely discussed and applied to describe the dynamics of stars and to construct models of
the evolution of the Universe.
\par
    Within a hydrodynamic approach, we will consider a version in which a quantum system of interacting
quarks and gluons is simulated classically by representing it as a perfect fluid having intrinsic degrees of
freedom, the spin and the non-Abelian color charge of particles forming the fluid. The classical nonrelativistic
chromohydrodynamic model of a fluid was proposed in \cite{GHK} without allowing for spin properties. In this study,
the relativistic variational theory of a perfect spin fluid with color charge in an external Yang--Mills field is
constructed in the Riemann-Cartan space $U_{4}$ with curvature and torsion. The approach used here, which is based on
the formalism of exterior forms, generalizes that from \cite{Am2}, where some simplifications were made, and that
from \cite{prep}, \cite{Ind}, where a different mathematical formalism was employed. The objective of this study is
to deduce the canonical energy-momentum tensor for the model of fluid under study. A mathematically correct derivation
of this tensor is desirable because it can be used as a basis for obtaining the classical equations of motion for a color
particle that generalize Wong's equations \cite{Won} to the case of the color $SU(3)$ group and which take into account
the particle spin \cite{prep}, \cite{Ind}. Concurrently, the model in which a hydrodynamic description of quark-gluon
plasma on the basis of the Landau model is generalized to include explicitly the structure of the QCD vacuum and the
color charge of quarks \cite{VP} can be substantiated with the aid of this energy-momentum tensor.

\section{DYNAMICAL VARIABLES \newline AND CONSTRAINTS}
\setcounter{equation}{0}

\par
    In the formalism of exterior forms, fields that are associated with 0-forms $\psi$ and with forms $\bar{\psi}$
conjugate to them and which are transformed according to the representation of the direct product of the Lorentz group
and the color $SU(3)$ group appear to be dynamical variables describing a fluid. An element of the fluid possesses
the velocity 4-vector $\vec{u}=u^{a}\vec{e}_{a}$, which can be used to construct the velocity 3-form
$u:=\vec{u}\rfloor\eta=u^{a}\eta_{a}$ and the velocity 1-form $*u=u_{a}\theta^{a}= g(\vec{u},\cdots)$ satisfying
the condition
\begin{equation}
*u \wedge u = -c^{2}\eta \;, \label{eq:4}
\end{equation}
which means that the squared velocity is $g(\vec{u},\vec{u}) = - c^{2}$, as it must. In the above expressions,
$\theta^{a}$ stands for basis 1-forms; $\eta$ is the volume 4-form; $\rfloor$ and $*$ denote, respectively, the
internal product and Hodge dualization operation; and the basis in the space of 3-forms $\eta_{b}$ and the basis
in the space 2-forms $\eta_{ab}$ are given by \cite{Tr1}
\begin{eqnarray}
&&\eta_{b} = \vec{e}_{b}\rfloor \eta = *\theta_{b} \; , \;\;\;\;\;\;\;\;\;\;\;\;\;
\theta^{a}\wedge \eta_{b} = \delta^{a}_{b}\eta \; ,  \\
&&\eta_{ab} = \vec{e}_{b}\rfloor \eta_{a} = *(\theta_{a}\wedge \theta_{b}) \; ,
\;\;\;\;\;
\theta^{c}\wedge \eta_{ab} = \delta^{c}_{b}\eta_{a} -
\delta^{c}_{a}\eta_{b}\; . \label{eq:3}
\end{eqnarray}
The basis $\vec{e}_{a}$ is assumed to be nonholonomic and orthogonal, and $g(\vec{e}_{a},\vec{e}_{b}) =: g_{ab} =
\mbox{diag}(1,1,1,-1)$.
\par
    The internal-energy density $\varepsilon$ in a fluid depends on extensive (additive) thermodynamic parameters -- the
particle-number density (concentration) $n$ and the entropy per particle $s$, as calculated in the reference frame
comoving with the fluid -- and on the quantities describing the intrinsic degrees of freedom of particles forming
the fluid, such as the particle spin tensor $S_{ab}$ and particle color charge $J_{m}$. The quantities characterizing
the fluid (including $\psi$ and $\bar{\psi}$) are taken to be averaged over an elementary 3-volume $V$ containing what
is referred to as a fluid element. The particle spin tensor and the particle color charge of the fluid under
study are given by
\begin{equation}
S_{ab} = \bar{\psi}M_{ab}\psi\;, \qquad J_{m} = \bar{\psi}I_{m}\psi\; ,\label{eq:SI}
\end{equation}
where $M_{ab}$ and $I_{m}$ are the generators of the corresponding representations of the Lorentz group and the
color $SU(3)$ group. On the other hand, a fluid element represents a statistical subsystem in which the number of
particles is so great that this subsystem can be treated as a quasiclosed system whose properties coincide with the
statistical properties of a fluid in a macroscopic state. For a fluid element, the first law of thermodynamics is
written as
\begin{equation}
d(\varepsilon V) = -p dV + Td(n s V) + \frac{1}{2}\omega^{ab}d(n S_{ab} V) +
\omega^{m}d(n J_{m}V)\; ,\label{eq:13a}
\end{equation}
where $p$ is the hydrodynamic pressure of the fluid. The factors $\omega^{ab}$ and $\omega^{m}$ characterize the
possible exchange of, respectively, spin and charge between fluid elements. Substituting the relation $V=N/n$, where $N$
is the total number of particles in a given fluid element, into (\ref{eq:13a}), we express the first law of
thermodynamics as \cite{prep}
\begin{equation}
d\varepsilon(n, s, S_{ab}, J_{m}) = \frac{\varepsilon + p}{n} dn +
n T ds + \frac{1}{2} n \omega^{ab}dS_{ab} + n \omega^{m}dJ_{m}\; ,
\label{eq:13}\end{equation}
where the coefficient of $dn$ has the meaning of a chemical potential. The above form of the chemical potential
indicates that the number of particles in the system is conserved. The chemical potential plays an important role
in describing the thermodynamic properties of quark-gluon plasma \cite{Kap}.
\par
    We assume that the fluid being considered moves in such a way that the laws of particle-number and entropy
conservation are satisfied. These conservation laws can be written as
\begin{eqnarray}
d(n u) = 0 \; , \qquad d(n s u) = 0 \; , \label{eq:7}
\end{eqnarray}
where the symbol $d$ denotes the operation of exterior differentiation. The first of these equations ensures
continuity of particle current lines in the fluid. The second equation expresses invariability of entropy along current
lines in the fluid. This is the second law of thermodynamics for an adiabatic flow of the fluid. As was indicated in
Frenkel's theory of a rotating electron \cite{Fren}, the spacelike nature of the spin tensor is a physical circumstance
of fundamental importance, which is expressed in fulfillment of the Frenkel condition $S_{ab} u^{b} = 0$. In terms of
exterior forms, this condition is represented as
\begin{equation}
(\vec{e}_{a}\rfloor {\cal  S})\wedge u = 0\; ,\qquad {\cal S} = \frac{1}{2}S_{ab}
\theta^{a}\wedge \theta^{b}\; , \label{eq:ufr}
\end{equation}
where we introduced the spin 2-form ${\cal S}$.

\section{LAGRANGIAN DENSITY AND \newline EQUATIONS OF MOTION OF A FLUID}
\setcounter{equation}{0}
\par
    In the formalism of exterior forms, the action functional is given by an integral of the Lagrangian density
4-form, which is taken to be
\begin{eqnarray}
&&{\cal L}_{fluid} = L_{fluid}\;\eta = -\varepsilon(n,s,\psi,\bar{\psi})\eta +
n\bar{\psi}{\cal D}\psi \wedge u - \chi n J_{m}{\cal F}^{m}\wedge *{\cal S} +
\nonumber \\
&& + n\lambda_{1}(*u\wedge u + c^{2}\eta) + n u\wedge  d\lambda_{2} +
n \lambda_{3} u\wedge ds + n \zeta^{a}(\vec{e}_{a}\rfloor {\cal S})\wedge u \;.
\label{eq:lag}
\end{eqnarray}
\par
    For independent variables in (\ref{eq:lag}) that describe the dynamics of the fluid in question,
we choose $n$, $s$, $\psi$, $\bar{\psi}$ and $u$. Constraints that are imposed on the independent variables and
which are specified by equations (\ref{eq:4}), (\ref{eq:7}) and (\ref{eq:ufr}), are taken into account via indefinite
Lagrange multipliers $\lambda_{1}$, $\lambda_{2}$,  $\lambda_{3}$ and $\zeta^{a}$. The second term in (\ref{eq:lag})
represents a kinetic term, in which the symbol ${\cal D}$ denotes the operation of exterior covariant
differentiation with respect to both gauge groups, the Lorentz group and the color $SU(3)$ group; that is,
\begin{equation}
{\cal D}\psi = d\psi + \frac{1}{2}\Gamma^{ab}M_{ab}\psi + A^{m}I_{m}\psi \;,
\label{eq:dif}
\end{equation}
where $\Gamma^{b}\!_{a}$ is the connection 1-form in the space $U_{4}$, and $A^{m}$ is the 1-form of the potential of
the gauge color field.
\par
    In (\ref{eq:lag}), the term that involves the coupling constant $\chi$ describes the possible spin-polarization
chromomagnetic effects. In this term, ${\cal  F}^{m}$ is the strength 2-form for the non-Abelian gauge color field;
that is,
\begin{equation}
{\cal F}^{m} = dA^{m} + \frac{1}{2}c_{p}\!^{m}\!_{q}A^{p}\wedge A^{q} =
\frac{1}{2}F^{m}\!_{ab}\theta^{a}\wedge \theta^{b}\; , \label{eq:F}
\end{equation}
where $c_{p}\!^{m}\!_{q}$ are the structure constants of the $SU(3)$ group.
\par
    To evaluate the variation of (\ref{eq:lag}), we will need the variations
\begin{eqnarray}
&&\eta \delta u^{a} = - \delta u \wedge \theta^{a} + \delta \theta^{a} \wedge u - u^{a}\delta \eta\; , \qquad
\delta \eta = \delta \theta^{a} \wedge \eta_{a} \; , \label{eq:17}\\
&&\delta *\!u = g_{ab}\theta^{a}\delta u^{b} + u^{b}g_{ab}\delta \theta^{a}\; , \label{eq:161}
\end{eqnarray}
which follow from the relations $\theta^{a}\wedge u = u^{a}\eta $ and $*u = g_{ab}u^{a}\theta^{b}$, respectively.
Equations (\ref{eq:17}) and (\ref{eq:161}) were obtained by considering that, in a nonholonomic orthogonal basis,
the components of the metric tensor $g_{ab}$ are constants (that is, they must not be varied).
\par
    The variation of (\ref{eq:lag}) with respect to the indefinite Lagrange multipliers yields the corresponding
constraint equations, whereas the variation with respect to the dynamical variables leads to variational equations
of motion for a perfect fluid having a non-Abelian color charge. We have
\begin{eqnarray}
\delta n : &&\quad (\varepsilon + p) \eta - n \bar{\psi}{\cal D}\psi\wedge u
+ \chi n J_{m}{\cal F}^{m}\wedge *{\cal S} - n u\wedge d\lambda_{2} = 0 \; ,\label{eq:21}\\
\delta s : &&\quad  T \eta + u \wedge d\lambda_{3} = 0 \;, \label{eq:22}\\
\delta u : &&\quad \bar{\psi}{\cal D}\psi - (-2\lambda_{1}*\!u + d\lambda_{2}
+ \lambda_{3} ds) + \zeta^{a}\vec{e}_{a}\rfloor{\cal S}= 0\; , \label{eq:25} \\
\delta \bar{\psi} : &&\quad \frac{\partial\varepsilon}{\partial\bar{\psi}}
\eta - n {\cal D}\psi\wedge u - n\zeta^{a} M_{ab}\psi \theta^{b}\wedge u + \nonumber \\
&&\quad + \frac{1}{2}\chi n J_{m}{\cal F}^{m}\wedge M_{ab}\eta^{ab}\psi +
\chi n J_{m}{\cal F}^{m}\wedge *{\cal S}\psi = 0\; , \label{eq:bpsi} \\
\delta \psi : &&\quad \frac{\partial\varepsilon}{\partial\psi}\eta
+ n {\cal D}\bar{\psi}\wedge u - n\zeta^{a} \bar{\psi}M_{ab}\theta^{b}\wedge u + \nonumber \\ &&\quad
+ \frac{1}{2}\chi n J_{m}{\cal F}^{m}\wedge \bar{\psi}M_{ab}\eta^{ab}
+ \chi n J_{m}{\cal F}^{m} \wedge *{\cal S}\bar{\psi} = 0\; , \label{eq:psi}
\end{eqnarray}
In the coordinate basis $\vec{e}_{\alpha} = \partial_{\alpha}$ equation (\ref{eq:22}) takes the form
$T = u^{\alpha}\partial_{\alpha}\lambda_{3}$, which unveils the physical meaning of the indeterminate Lagrange
multiplier $\lambda_{3}$ as a "thermasion" (see \cite{Sh}).
\par
    In evaluating the derivatives of the internal energy $\varepsilon$ with respect to $\psi$ and $\bar{\psi}$,
we must consider in equations (\ref{eq:psi}) and (\ref{eq:bpsi}) that, by virtue of (\ref{eq:13}), the internal
energy depends on $\psi$ and $\bar{\psi}$ only through its dependence on $S_{ab}$ and $J_{m}$:
\begin{equation}
\varepsilon(n, s, \bar{\psi}, \psi) = \varepsilon(n, s,\bar{\psi}M_{ab}\psi,
\bar{\psi}I_{m}\psi )\; . \label{eq:E}
\end{equation}
Therefore, we have
\begin{eqnarray}
&&\frac{\partial\varepsilon}{\partial\psi} = \frac{1}{2}n \omega^{ab}\bar
{\psi}M_{ab} + n \omega^{m}\bar{\psi}I_{m}\; , \label{eq:E1}\\
&&\frac{\partial\varepsilon}{\partial\bar{\psi}} = \frac{1}{2}n \omega^{ab}
M_{ab}\psi + n \omega^{m}I_{m}\psi \; .\label{eq:E2}
\end{eqnarray}
\par
    Multiplying equation (\ref{eq:25}) by $u$ from the right in such a way as to obtain the corresponding
exterior product and using (\ref{eq:21}), we find that the Lagrange multiplier $\lambda_{1}$ can determined from
the relation
\begin{equation}
2 n c^{2} \lambda_{1}\eta = (\varepsilon + p)\eta + \frac{1}{2} \chi n
J_{m}{\cal F}^{m}\wedge *{\cal S}\; . \label{eq:28}
\end{equation}
\par
    With the aid of equation (\ref{eq:21}) and constraints (\ref{eq:4}), (\ref{eq:7}) and (\ref{eq:ufr}), it can
easily be shown that the Lagrangian density 4-form (\ref{eq:lag}) is proportional to the hydrodynamic pressure of
the fluid:
\begin{equation}
{\cal L}_{fluid} = p \eta \; . \label{eq:p}
\end{equation}
This means that there is a correct limiting transition from the variational theory of a perfect spin fluid with
intrinsic color charge to the variational theory of a conventional perfect fluid.
\par
    To determine the Lagrange multipliers $\zeta^{a}$, we must evaluate the expression $u^{b}{\cal D}(nS_{ab} u)$
by using the equations of motion of the fluid, which are given by (\ref{eq:bpsi}) and (\ref{eq:psi}). As a result,
we obtain
\begin{equation}
\zeta^{c}S_{ca} = \frac{1}{c^{2}}\dot{S}_{ab}u^{b} - \frac{1}{c^{2}}  S_{ac}
u_{b}(\chi J_{m}F^{mbc} + \omega^{bc})\; .\label{eq:z}
\end{equation}
where the dot over the variable $S_{ab}$ denotes the operation of differentiation. For an arbitrary tensor object
$\Phi$, this operation is defined as
\begin{equation}
\dot{\Phi}^{a}\!_{b} := *\!(u\wedge {\cal D}\Phi^{a}\!_{b})\; .\label{eq:27}
\end{equation}
\par
    Using expression (\ref{eq:z}) and the identities ${\cal D}M_{ab} = 0$, we find that, for particles of the fluid
being considered, the variation of the spin tensor is governed by the law
\begin{eqnarray}
&&u\wedge {\cal D}S_{ab} + \frac{2}{c^{2}}S_{[a}\!^{c} u_{b]}\dot{u}_{c}\eta =
\nonumber \\
&& = - 2 S_{[a}\!^{c}(\chi \eta_{b]c}\wedge {\cal F}^{m}J_{m} +
\omega_{b]c}\eta) + \frac{2}{c^{2}}S_{[a}\!^{c} u_{b]} u^{d}
(\chi F^{m}\!_{cd}J_{m} + \omega_{cd})\eta \; . \label{eq:sp}
\end{eqnarray}
If right-hand side of this equation is set to zero, we arrive at the law that controls the variation of the spin
tensor in the Weyssenhoff--Raabe perfect spin fluid \cite{Ray-Sm}--\cite{Bab:iz}. Equation (\ref{eq:sp}) demonstrates
that spin non-conservation along current lines in the fluid is associated with spin-chromomagnetic interaction and
with the possible exchange of spin between fluid elements.

\section{EQUATIONS FOR A NON-ABELIAN\newline GAUGE COLOR FIELD}
\setcounter{equation}{0}
\par
    To obtain equations for a non-Abelian gauge color field, the Lagrangian density 4-form (\ref{eq:lag}) must be
supplemented with the Lagrangian density 4-form describing the color field. Accordingly, we have
\begin{equation}
{\cal L}_{matter} = {\cal L}_{fluid} + {\cal L}_{field} \;, \label{eq:Lm}
\end{equation}
where
\begin{equation}
{\cal L}_{field} = -\frac{\alpha}{2}{\cal F}^{m}\wedge *{\cal F}_{m} =
-\frac{\alpha}{4}F^{m}\!_{ab}F_{m}\!^{ab}\eta\;, \label{eq:Lfd}
\end{equation}
$\alpha$ being the coupling constant. The indices of the type $m$ on the gauge field are transformed according to the
adjoint representation of the color $SU(3)$ group. These indices are raised and lowered with the aid of the metric
tensor $g_{mn} = - \frac{1}{2}c_{m}\!^{p}\!_{q} c_{n}\!^{q}\!_{p}$.
\par
    The field equations are obtained by varying (\ref{eq:Lm}) with respect to the 1-form of the potential $A^{m}$ for
the gauge color field. In this way, we obtain
\begin{eqnarray}
\delta A^{m}: &&\quad {\cal D}(\alpha *\!{\cal F}_{m} + \chi n J_{m}*\!
{\cal S}) = n J_{m} u \; . \label{eq:A} \end{eqnarray}
This equation must be supplemented with the Bianchi identity  ${\cal D}{\cal F}^{m} = 0$ for the non-Abelian gauge field.
\par
    Using the equations of motion of the fluid, which are given by (\ref{eq:bpsi}) and (\ref{eq:psi}), and taking into
account the identity ${\cal  D}I_{m} = 0$, we find that evolution of the non-Abelian color charge is governed by the
law
\begin{equation}
u\wedge {\cal D}J_{m} = - c_{m}\!^{p}\!_{n} J_{p} (\chi {\cal F}^{n}\wedge
*{\cal S} + \omega^{n} \eta )\; . \label{eq:J}
\end{equation}
This equation shows that, in the non-Abelian case, spin-chromomagnetic interaction may result in color-charge
nonconservation along current lines in the fluid. It is interesting to note that, in the Abelian case, the electric
charge is always conserved along current lines in the fluid by virtue of the condition $c_{m}\!^{p}\!_{n} = 0$,
which holds for Abelian fields.
\par
    The classical equations of motion that were obtained here for a non-Abelian color field and which take into account
spin and color degrees of freedom represent a self-consistent set. This circumstance is of particular interest because
the structure of the QCD vacuum may greatly depend on QCD fields other than the gluon field. This suggests that, even at
the classical level, it is advisable to solve model problems with allowance for currents of charged particles having
various origins \cite{BVK}. Indeed, we see that, in equations (\ref{eq:A}) and (\ref{eq:J}), the current of the source
is determined both by the current of the particle color charge, the corresponding term appearing on the right-hand side,
and by the spin correction appearing on the left-hand side of equation (\ref{eq:A}). This makes it possible to obtain
deeper insight into the global dynamics of the system.

\section{ENERGY-MOMENTUM TENSOR \newline OF A PERFECT SPIN FLUID\newline WITH A COLOR CHARGE}
\setcounter{equation}{0}
\par
    According to the general concepts of gauge field theory in Riemann--Cartan space \cite{Kib}, \cite{Fr1},\-\cite{Tr1},
the geometric properties of spacetime are determined by the canonical energy-momentum tensor and by the matter spin tensor.
The spin 3-form $\Delta^{a}\!_{b}$ defined in terms of the variational derivative of the Lagrangian density
4-form (\ref{eq:Lm}) with respect to the connection 1-form $\Gamma^{b}\!_{a}$ in the Riemann--Cartan space $U_{4}$ as
\begin{equation}
\Delta\!^{a}\!_{b} := - \frac{\delta{\cal L}_{matter}}{\delta\Gamma^{b}\!_{a}} =
\frac{1}{2} n S^{a}\!_{b} u \;  \label{eq:38}
\end{equation}
plays the role of the matter spin tensor in the formalism of exterior forms. Within this framework, the canonical
energy-momentum tensor is taken to be the energy-momentum 3-form $\Sigma_{a}$ defined as the variational derivative
of the Lagrangian density 4-form (\ref{eq:Lm}) with respect to the basis 1-form $\theta^{a}$. In the space $U_{4}$,
variation with respect to $\theta^{a}$ is independent of the variation with respect to $\Gamma^{b}\!_{a}$; in a
nonholonomic orthogonal basis, the condition that ensures the consistency of the metric and connection and which
represents the constraint ${\cal  D}g_{ab}=0$ on the independent variables holds by virtue of antisymmetry of the
connection 1-form, $\Gamma^{ab} = -\Gamma^{ba}$, so that additional terms with indeterminate Lagrange multipliers
are not needed for this purpose.
\par
    Evaluating $\Sigma_{a}$ for the Lagrangian density (\ref{eq:Lm}) with the aid of (\ref{eq:25})--(\ref{eq:psi}),
(\ref{eq:28}) and (\ref{eq:z}), we obtain
\begin{eqnarray}
\Sigma_{a} &=& - \frac{\delta{\cal L}_{matter}}{\delta\theta^{a}} =
\Sigma^{fluid}_{a} + \Sigma^{field}_{a}\; ,\\
\Sigma^{fluid}_{a} &=& p\Pi^{b}_{a}
\eta_{b} + \frac{1}{c^{2}}n(\varepsilon^{*}\delta^{b}_{a}
+ \dot{S}_{a}\!^{b}) u_{b}u - \chi n J_{m}F^{m}\!_{ab}S^{bc}\eta_{c} -
\nonumber \\
&& -\frac{1}{c^{2}}\chi nS_{a}\!^{c}(J_{m}F^{m}\!_{bc} -\omega_{bc})u^{b}u
\; , \label{eq:fl}\\
\Sigma^{field}_{a} &=& \alpha (F^{m}\!_{ac}F_{m}\!^{bc} - \frac{1}{4}
\delta^{b}_{a}F^{m}\!_{cd}F_{m}\!^{cd})\eta_{b} = \nonumber \\
&& = \frac{\alpha}{2}\left ((e_{a}\rfloor {\cal F}^{m}) \wedge *{\cal F}_{m}
- {\cal F}^{m}\wedge (e_{a}\rfloor *\!{\cal F}_{m})\right )\; .\label{eq:fd}
\end{eqnarray}
Here $\Pi^{b}_{a}$ denotes the components of the projector tensor

\begin{equation}
\Pi^{b}_{a} = \delta^{b}_{a} + \frac{1}{c^{2}}u^{b}u_{a}\; ,\label{eq:Pi}
\end{equation}
and $\varepsilon^{*}$ is the effective energy per particle \cite{prep} given by
\begin{equation}
\varepsilon^{*}\eta = \varepsilon_{0}\eta + \chi J_{m}{\cal F}^{m}\wedge
*{\cal S}\; , \qquad \varepsilon_{0} = \frac{\varepsilon}{n}\; . \label{eq:ef}
\end{equation}
Expression (\ref{eq:ef}) generalizes the result from \cite{Cor}, which was obtained in the presence of an electromagnetic
field, to the non-Abelian case.
\par
    This expression can easily be explained in quantum theory. The second term in (\ref{eq:ef}) describes the
interaction of spin and color charge with an external color field. To formulate this point more precisely, we indicate
that, depending on helicity, particles in a definite color state interact with the corresponding state of the field;
as a result, the effective energy increases (or decreases). On the other hand, particles that possess the same helicity,
but which occur in a different color state do not interact with this field state \cite{BVK}. In the Abelian case (QED),
there is the analogous effect of an increase in the electron mass owing to its interaction with  an electromagnetic
field \cite{BLP}.
\par
    Let us introduce the dynamical momentum of a fluid element per particle via the relation
\begin{equation}
\pi_{a}\eta = -\frac{1}{nc^{2}}*\!u\wedge \Sigma^{fluid}_{a}\;, \quad
\pi_{a} = \frac{1}{c^{2}}\varepsilon^{*}u_{a} - \frac{1}{c^{2}}S_{a}\!^{c}
\left (\dot{u}_{c} + u^{b}(\chi J_{m}F^{m}\!_{bc} + \omega_{bc})\right ) \; .
\label{eq:pi}\end{equation}
The canonical energy-momentum 3-form (\ref{eq:fl}) can then be represented as
\begin{equation}
\Sigma^{fluid}_{a} = p\eta_{a} + n\left (\pi_{a} +\frac{p}{nc^{2}}
u_{a}\right ) u + \chi n(\vec{e}_{a}\rfloor {\cal F}^{m}J_{m})\wedge *{\cal S}\; . \label{eq:sig}
\end{equation}
This form is more convenient for applications. We can see that the hydrodynamic pressure $p$ of the fluid manifests
itself in two ways: it impedes the compression of the fluid [first term in (\ref{eq:sig})], on one hand, and
effectively increases the mass of a fluid element [second term in (\ref{eq:sig})], on the other hand. In the gravitational
collapse of astrophysical objects simulated by quark-gluon plasma, the second manifestation of the pressure is dominant;
therefore, the growth of the pressure cannot  prevent a collapse.
\par
    Note that the full energy-momentum 3-form for a perfect spin fluid with a color charge in an external color field
can also be represented in the form
\begin{eqnarray}
\Sigma_{a} &=& \vec{e}_{a}\rfloor \left (p\eta - \frac{\alpha}{2}{\cal F}^{m}\wedge
*{\cal F}_{m}\right ) + n\left (\pi_{a} +\frac{p}{nc^{2}}u_{a}
\right ) u + \nonumber \\
&& + (\vec{e}_{a}\rfloor {\cal F}^{m})\wedge (\alpha *\!{\cal F}_{m} +
\chi nJ_{m}*\!{\cal S})\; , \label{eq:st}
\end{eqnarray}
which proves to be useful in some cases.

\section{CONCLUSION}
\par
    On the basis of the variational approach, we have given a consistent derivation of the classical energy-momentum
tensor for a perfect fluid having intrinsic spin and color degrees of freedom and interacting with a non-Abelian color
gauge field. In the future, a self-consistent set of the dynamical equations of motion of a classical particle with
spin and with a non-Abelian color charge in a color field will be obtained against the background of Riemann--Cartan
geometry with curvature and torsion by using the above energy-momentum tensor as a starting point. These equations,
together with the continuity condition for current lines in the fluid, the equation of state of the fluid, and
the equations for the color field, will represent the full set of equations necessary for describing quark-gluon plasma.
\par
    By taking into account the variational theory of spin-dilaton fluid \cite{Los}, the theory proposed here can be
generalized to the case in which particles forming the fluid possess not only a spin and a color charge but also a dilaton
charge. Presumably, analysis of such objects will be useful in QCD in connection with intensive investigations of the
O++ glueball (dilaton) \cite{Ag} and with the possibility of studying, on this basis, a nonperturbative gluon condensate
up to the point of a phase transition between hadronic matter and quark-gluon plasma.


\begin{thebibliography}{50}
\bibitem{Sed}
Sedov L.I., {\it A Course of Continuum Mechanics}, vols. 1--4, Groningen: Wolters-Noordhoff, 1971-1972.
\bibitem{Pol}
Polyakov A.M., {\it Kalibrovochnye polya i struny} (Gauge Fields and Strings), Chernogolovka: Inst. Teor. Fiz. im.
Landau, 1995.
\bibitem{Jac}
Jackiw R., {\it hep-th/9602122}, 1996.
\bibitem{Wil}
Williams L.B., {\it The Origins of Field Theory}, New York: Random House, 1996.
\bibitem{Mon}
Monastyrskii M.I., {\it Topologiya kalibrovochnykh polei i kondensirovannykh sred} (Topology of Gauge
Fields and Condensed Media), Moscow: PAIMS, 1995.
\bibitem{VP}
Vshivtsev A.S. and Peregudov D.V., Yad. Fiz., 1997, vol. 60, p. 1481 [Phys. At. Nucl. (Engl. Transl.), vol. 60, p. 1344].
\bibitem{GHK}
Gibbons J., Holm D.D., and Kupershmidt B.A., {\it Physica D: Nonlinear Phenomena}, 1983, vol. 6, p. 179;
Holm D.D. and Kupershmidt B.A., {\it Phys. Rev. D: Part. Fields}, 1984, vol. 30, p. 2557.
\bibitem{Am2}
Amorim R., Phys. {\it Rev. D: Part. Fields}, 1986, vol. 33, p. 2796.
\bibitem{prep}
Bagrov V.G., Babourova O.V., Vshivtsev A.S., and Frolov B.N, {\it Preprint of Tomsk Branch,
Siberian Division, USSRAcad. Sci.},
1988, no. 33.
\bibitem{Ind}
Babourova O.V. and Frolov B.N., {\it Mahavisva} (India), 1991, vol. 4, nos. 1/2, p. 55.
\bibitem{Won}
Wong S.K., {\it Nuovo Cimento}, 1970, vol. 65, p. 689.
\bibitem{Tr1}
Trainman A., {\it Symp. Math.}, 1973, vol. 12, p. 139.
\bibitem{Kap}
Kapusta J., {\it Finite Temperature Field Theory}, New York: Cambridge Univ. Press, 1989.
\bibitem{Fren}
Frenkel J., {\it Z. Phys.}, 1926, vol. 37, p. 243; Frenkel, Ya.L, {\it Sobranie izbrannykh trudov v dvukh tomakh}
(Collected Works in Two Volumes), Moscow: Akad. Nauk SSSR, 1958, vol. 2, p. 460.
\bibitem{Sh}
Schutz B.F., {\it Phys. Rev. D: Part. Fields}, 1970, vol. 2, p. 2762.
\bibitem{Ray-Sm}
Ray J.R. and Smally L.L., {\it Phys. Rev. D: Part. Fields}, 1983, vol. 27, p. 1383.
\bibitem{Rit1}
de Ritis R., Lavorgna M., Platania G., and Stornaiolo C., {\it Phys. Rev. D: Pan. Fields}, 1983, vol. 28, p. 713.
\bibitem{Ob-Kor}
Obukhov Yn.N. and Korotky V.A., {\it Classical Quantum Gravity}, 1987, vol. 4, p. 1633.
\bibitem{Bab:iz}
Babourova O.V., {\it Izv. vyssh. Uchebn. Zaved., Fiz.}, 1989, no. 10, p. 101.
\bibitem{BVK}
Bagrov V.G., Vshivtsev A.S., and Ketov S.V., {\it Dopolnitelnye glavy matematicheskoi fiziki (kalibro-vochnye polya)}
(Supplementary Chapters of Mathematical Physics: Gauge Fields), Tomsk: Tomsk Gos. Univ., 1990.
\bibitem{Kib}
Kibble T.W.B., {\it J. Math. Phys.}, 1961, vol. 2, p. 212.
\bibitem{Fr1}
Frolov B.N., {\it Vestn. Mask. Univ., Ser. 3: Fiz. Astron.}, 1963, no. 6, p. 48.
\bibitem{Cor}
Corben H.C., {\it Classical and Quantum Theory of Spinning Particles}, San Francisco: Holden-Day, 1968.
\bibitem{BLP}
Berestetskii V.B., Lifshitz E.M., and Pitaevskii L.P., {\it Relativistic Quantum Theory}, Oxford: Pergamon, 1971.
\bibitem{Los}
Babourova O.V. and Frolov B.N, {\it gr-qc/9612055}.
\bibitem{Ag}
Agasyan N.O., {\it Pis'ma Zh.  Eksp.  Tear.  Fiz.},  1993, vol. 57, p. 208.
\end{thebibliography}
\end{document}